\documentstyle[11pt,newpasp,twoside,epsf]{article}

\begin{document}

\title{Interacting Stellar Wind and Photoionization Models of the SN
1987A remnant}

\author{Robert Link, Duane L. Rosenberg, and Roger Chevalier}
\affil{Department of Astronomy, University of Virginia, PO Box 3818,
Charlottesville, VA, 22903-0818}

\begin{abstract}
We are investigating the SN 1987A remnant by modeling the
circumstellar environment of the progenitor star.  Interacting stellar
winds models have been reasonably successful at reproducing the gross
features of the nebula, but some details, such as the early rise of the
radio and x-ray emission from the
supernova blast and the presence of the outer rings, are not explained
in a pure wind model.  
In this paper we
describe preliminary results from 2-D models that incorporate the
effects of photoionization due to radiation from the central star.
These models have successfully produced a thick HII region, as needed
to produce the early radio and X-ray emission.
The HII region is thickest away from the equatorial plane.  The models
have also produced a feature resembling the outer rings, but we
suspect this to be an artifact of the 2-D calculations that would not
persist in 3-D.  
\end{abstract}

\section{Introduction}
\label{sec:intro}

\begin{figure}
\plottwo{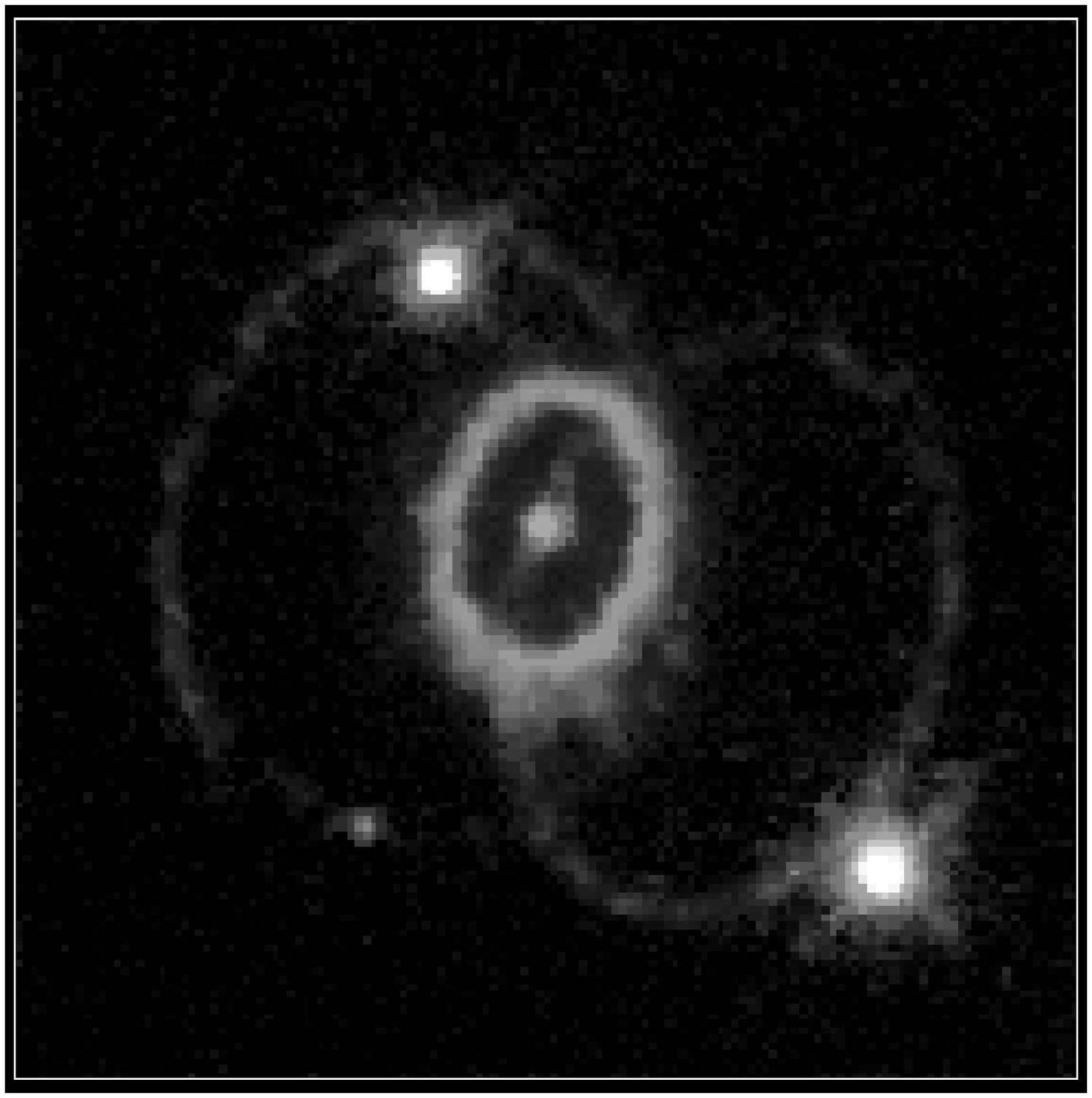}{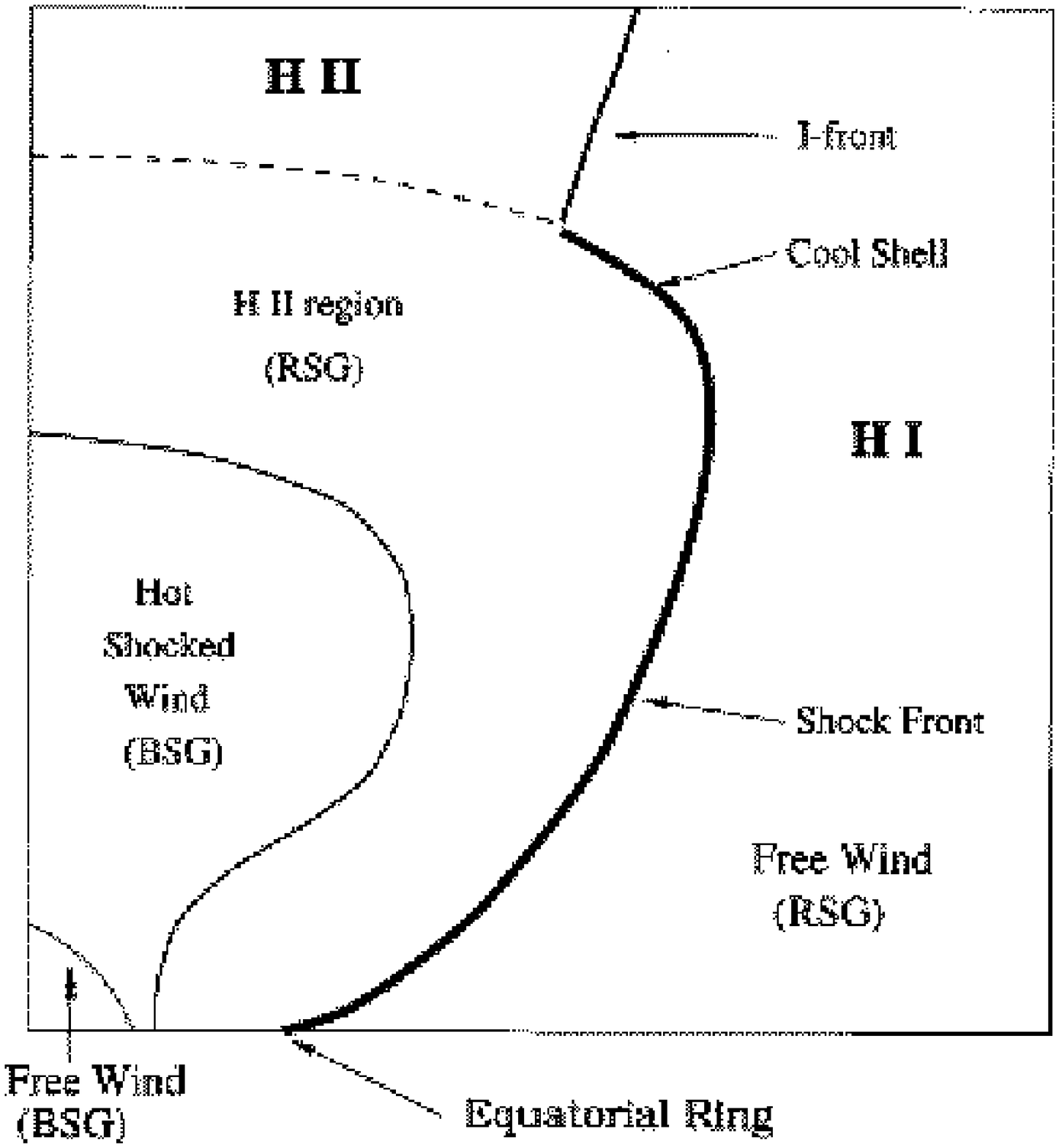}
\caption{Left: HST image of the SN1987A nebula (Burrows et
al. 1995). Right: schematic diagram of the structure of the nebula
(Chevalier \& Dwarkadas 1995).  The schematic shows a two-dimensional
section containing the symmetry axis.}
\label{fig:HST87a}
\end{figure}

Observations of the SN 1987A nebula give us the clues that we will
need to model its structure.  Figure~\ref{fig:HST87a} shows an HST
image of the nebula (Burrows et al. 1995) and a schematic diagram of
the nebular structure.  The HST image shows two of the salient
features of the nebula: the inner and outer rings.  The diagram shows
a slice containing the symmetry axis.  The inner ring lies in the
equatorial plane at a radius of about $0.6$ lt yr and is expanding with
a velocity of 10.3 km s$^{-1}$ (Crotts \& Heathcote 1991), while the outer
rings are at an angle of about 55$^{\circ}$ from the equatorial plane
and an axial radius of about $1.4$ lt yr.  A tenuous shell connects the
inner and outer rings (Crotts, Kunkel, \& Heathcote. 1995), but the
shell does not extend past the outer rings; that is, the nebula is
open along the symmetry axis.

In an interacting winds model for the nebula, a fast, hot wind ejected
during a blue supergiant phase overtakes a slow, dense wind ejected
from a previous red supergiant phase.  In order to produce the
equatorial ring the RSG wind must be asymmetrical, with higher density
at the equator and lower density at the symmetry axis.  Blondin \&
Lundqvist (1993) performed numerical simulations of this interacting
winds scenario for adiabatic and radiatively cooling gas.  These
simulations showed that the interacting winds model could produce the
inner ring and the overall shape of the nebula.  However, the Blondin
and Lundqvist models did not produce the outer rings, and the shell in
these models is complete all the way to the symmetry axis.  Therefore,
the interacting wind models need some refinement if they are to
describe the SN 1987A nebula.

Radio and x-ray observations of the nebula since 1990 provide
additional information.  These observations indicate that the
supernova blast has encountered a region of density intermediate to
the inner and outer winds (Gorenstein, Hughes, \& Tucker 1994;
Beuermann, Brandt, \& Pietsch 1994; Staveley-Smith et al. 1993;
Gaensler et al. 1997).  The
slow expansion of the radio source indicates that this material is too
dense to be shocked BSG wind.  Chevalier \& Dwarkadas (1995) proposed
that the intermediate density material is RSG wind that was
photoionized by radiation from the progenitor, a B3 Ia star.

In this paper we explore the interacting winds scenario, including
photoionization from the progenitor star, as a model for the SN 1987A
nebula.  Section~\ref{sec:numerical} discusses the features of our
numerical fluid dynamics code.  Section~\ref{sec:rslts} gives some
preliminary results from our numerical simulations, and
section~\ref{sec:conclusion} compares the results to SN 1987A
observations and discusses our plans for future refinements to the
models.

\section{Numerical methods}
\label{sec:numerical}

For our calculations we use a two-dimensional Eulerian finite
difference code with a van Leer monotonic transport algorithm.  We use
an ideal gas equation of state, $p=(\gamma-1)\epsilon$, where
$\epsilon$ is the internal energy density, and the adiabatic
constant, $\gamma = 5/3$.  The coordinate system is spherical-polar
with reflecting boundary conditions at the equator and pole.  The
inner boundary condition is an inflow condition, set to match the
characteristics of the BSG wind, and similarly the outer boundary
condition is an outflow condition set to the characteristics of the
RSG wind.  

In addition to the standard fluid dynamics equations, the code has
several features that are necessary for this particular problem.
These features include grid expansion, ionization and recombination,
radiative cooling, and separate advection for neutral and ionized gas.
They are described fully in Rosenberg, Link, \& Chevalier  (2000).

The RSG wind is
characterized by its the mass loss rate, velocity, and temperature,
$\dot{M}_R$, $v_R$, and $T_R$.  Since the RSG wind is asymmetric,
$\dot{M}_R$ is actually the mass loss rate that would be experienced
if the rate were everywhere equal to its value at the equatorial
plane.  The angular dependence of the mass loss rate is described by 
\begin{equation}
\dot{M}_R(\theta) = \dot{M}_R\left(\frac{\pi}{2}\right) \left[1 - A
\frac{\exp(-2\beta \cos^2(\theta)) - 1}{\exp(-2\beta) - 1}\right],
\end{equation}
where $A = 1-1/R$, $R$ is the ratio of equatorial to polar density in
the wind, and $\beta$ controls the steepness of the falloff from the
equatorial value.  Larger values of $\beta$ produce distributions more
strongly peaked at the equator; while smaller values produce a more
gradual falloff.  The BSG wind is similarly described by its mass loss
rate, velocity, and temperature, $\dot{M}_B$, $v_B$, and $T_B$.
Finally, the ionizing flux is described by the total flux of photons
above the ionization threshold, $S_*$.
The temperatures $T_A$ and $T_B$ are sufficiently low that they do not
affect the results.
Compared to the computations of Blondin \& Lundqvist (1993), 
$S_*$
is the only additional parameter.

Table~\ref{tbl:models} gives the parameters for the models used in our
calculations.  Model `A' is a fiducial model that was chosen
for comparison with previous results (\emph{e.g.} Blondin \& Lundqvist
1993).  

\begin{table}
\caption{Model Parameters}
\label{tbl:models}
\begin{center}
\begin{tabular}{rlllll}
\tableline
Identifier & $\dot{M}_R$ ($10^{-5} M_{\sun}$/yr) & $v_R$ (km/s) 
 & $R$ & $\beta$ & $S_*$ ($10^{45}$ s$^{-1}$) \\
           & $\dot{M}_B$ ($10^{-8} M_{\sun}$/yr) & $v_B$ (km/s) \\
\tableline
A & 7.5 & 5.0  & 20 & 4 & 4.0 \\
  & 7.5 & 450   \\
XA& 7.5 & 5.0  & 20 & 4 & 0.0 \\
  & 7.5 & 450   \\
B & 7.5 & 5.0  & 20 & 8 & 4.0 \\
  & 30  & 450   \\
C & 4.0 & 5.0  & 20 & 8 & 4.0 \\
  & 7.5 & 450   \\\tableline\tableline
\end{tabular}
\end{center}
\end{table}

\section{Results}
\label{sec:rslts}

\begin{figure}
\plottwo{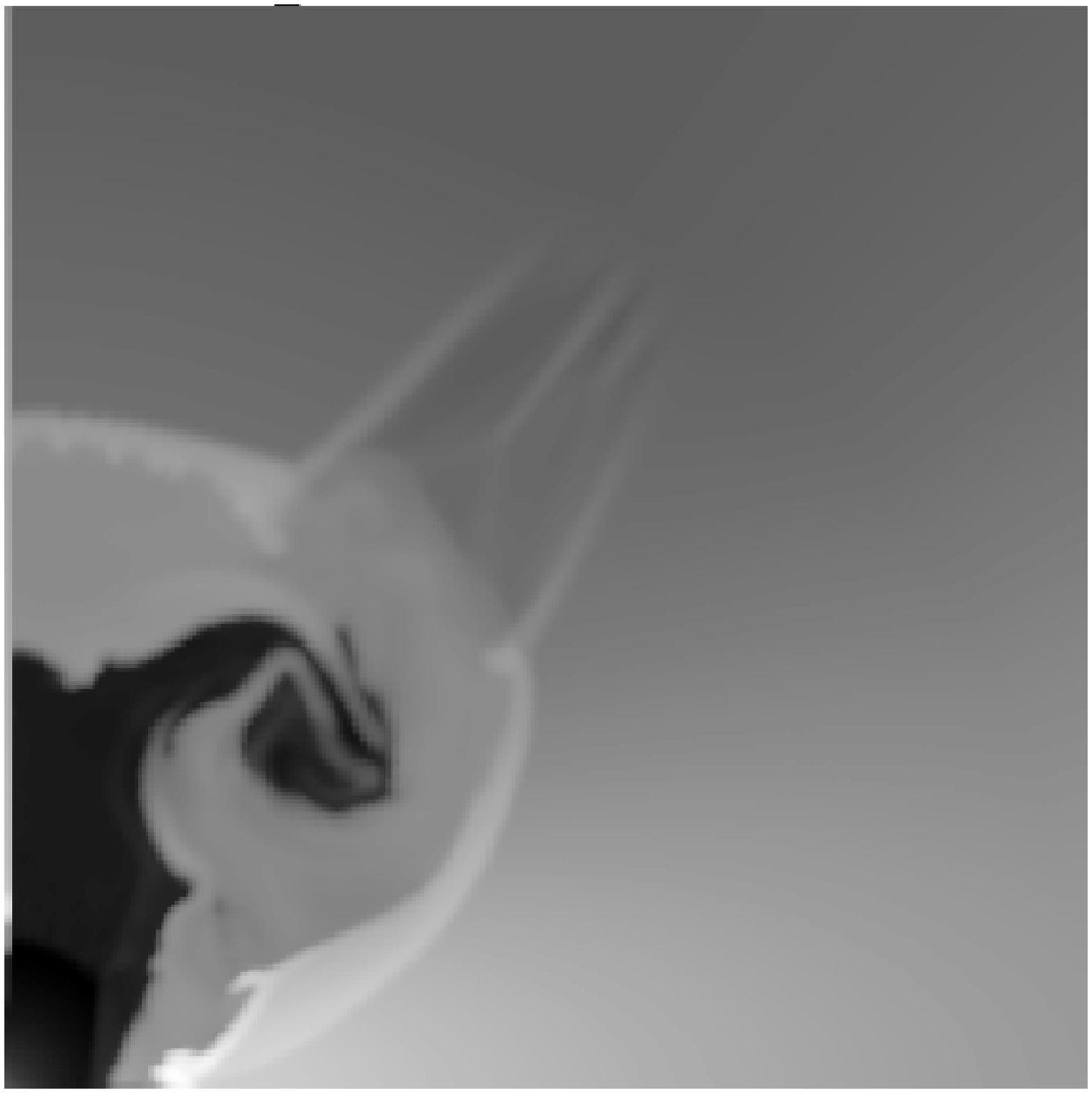}{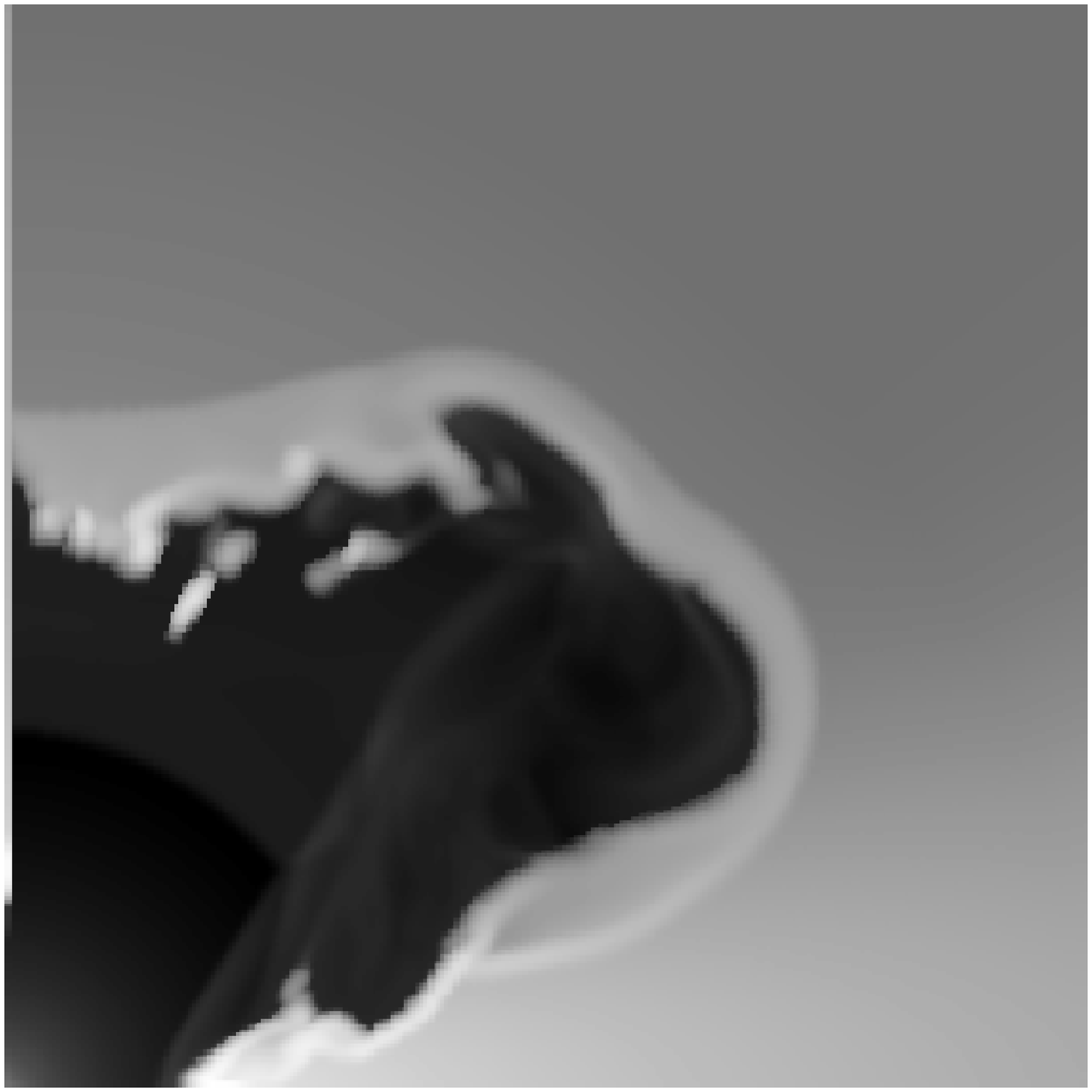}
\caption{The final states of models `A' (left) and `XA' (right).
Higher density is represented by lighter colors.  The two models are
identical, except that model `XA' has no photoionizing flux.  The
medium gray region just inside the neutral shell in model `A' is
composed of photoionized RSG wind.}
\label{fig:fiducial}
\label{fig:xa}
\end{figure}

Figure~\ref{fig:fiducial} shows a greyscale image of the final state
of the fiducial calculation.  The criterion for stopping the
calculation is that the radius of the inner ring in the calculation
reaches $6\times 10^{17}$ cm, which is approximately the measured
radius of the inner ring in the nebula.  The figure shows clearly the
inner ring and HII region.  However, there is no clear outer ring.
Moreover, although the ionization front has broken out at the middle
latitudes, the neutral shell is still intact near the pole, in
contrast to observations.
Further evolution of the model would result in photoionization of
the polar shell.

Model `XA' (fig.~\ref{fig:xa}) has identical parameters to model `A',
except that the photoionizing flux has been turned off.  Since there
is no photoionizing flux, there is no HII region.  The properties of
the inner ring, including the expansion velocity, are mostly unchanged
from the photoionization case; however, the structure in the middle
latitudes is quite different.  Particularly striking is the
fragmentation in the neutral shell, which leads to dense clumps that
look like they might form outer rings; however, these clumps appear to
be formed by a Rayleigh-Taylor or similar instability, and therefore
they should not be expected to exhibit the azimuthal symmetry required
to form the outer rings.

\begin{figure}
\plottwo{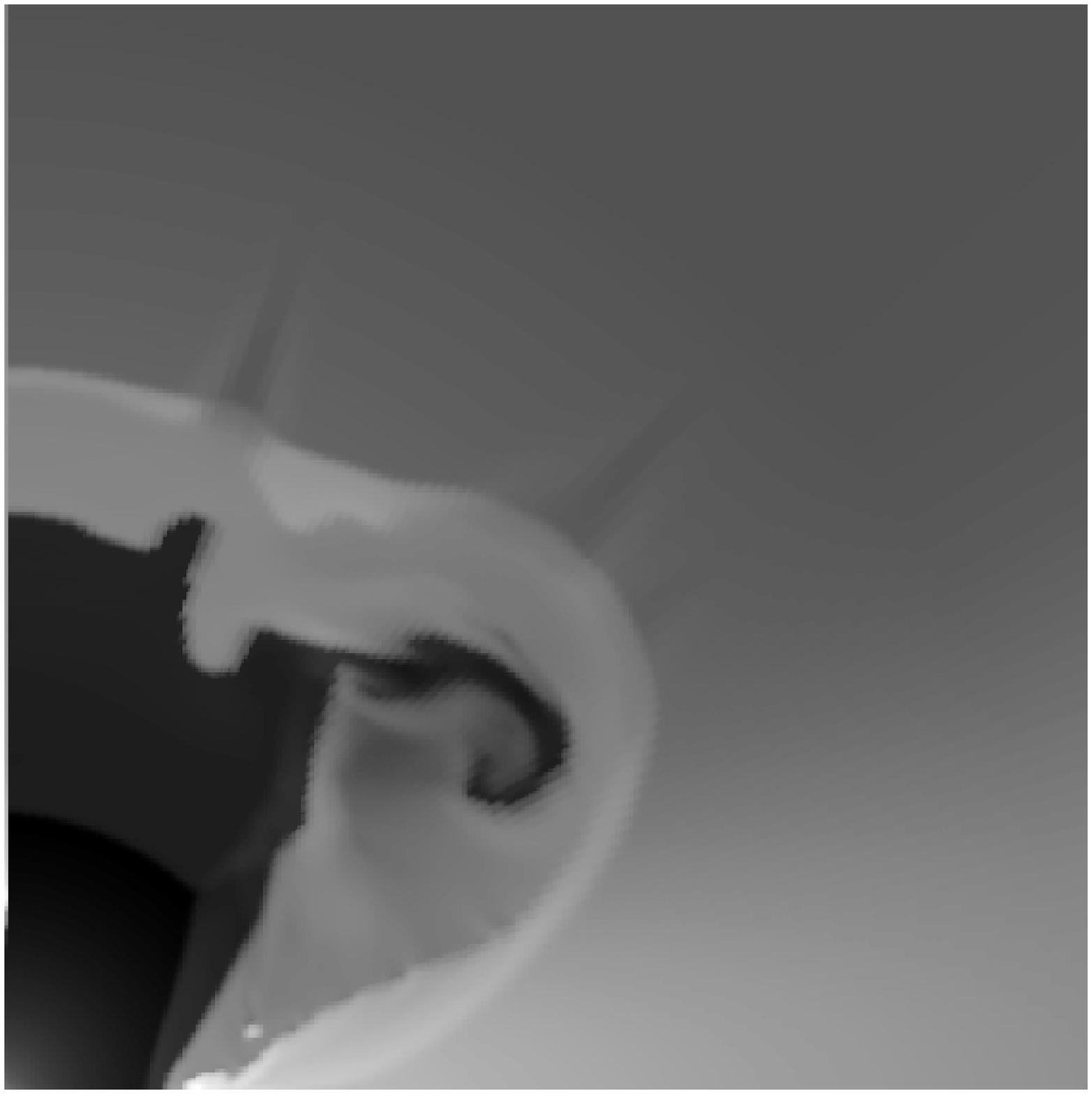}{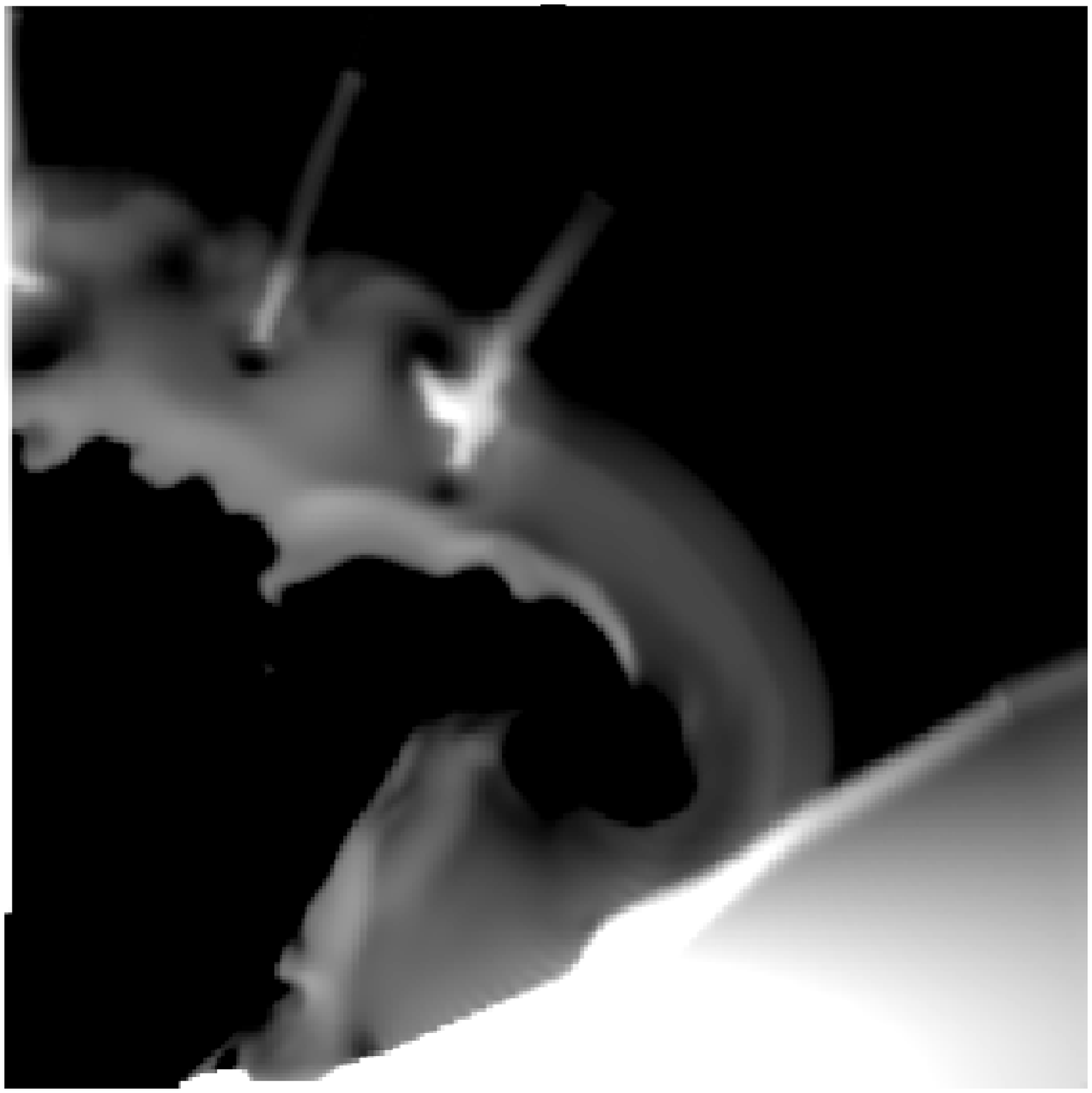}
\caption{The final state of models `B' (left) and `C' (right). Model
`C' forms a `ring' by ionizing most of the outer shell, leaving a blob
of neutral gas.}
\label{fig:b}
\end{figure}

The left panel of figure~\ref{fig:b} shows the final state of Model
`B'.  Model `B' has a stronger BSG wind than the fiducial model, and
it also has a more strongly peaked (higher $\beta$) RSG wind.  We see
in figure~\ref{fig:b} the beginning of a high-latitude ionization
front breakout that could eventually destroy the neutral shell near
the pole.  There is also a neutral clump that might form an outer
ring; however, a mid-latitude ionization front breakout has
disconnected the clump from the rest of the neutral shell.  The HII
region is thick throughout most of the model; however, right at the
equatorial plane it becomes very thin.  Apparently the ionized gas is
advected away from the equator, possibly due to the strong peak in the
density of the RSG wind.

Finally, Model `C' is shown in the right panel of figure~\ref{fig:b}.
The density of the RSG wind is lower than in the other simulations we
have discussed; consequently, most of the RSG wind above about
$30^{\circ}$ is ionized.  A dense blob of neutral gas remains at
mid-high latitude.  This neutral gas appears to survive longer than
the rest of the shell because it is compressed by converging shocks
preceding the ionization fronts.  It is possible (although not
certain) that such a blob could form a ring in three dimensions;
however, it does not seem possible to form both a ring and a
connecting shell through this mechanism.  Moreover, the `ring' is
short-lived.  The photoionizing flux must last long enough to blow
away the surrounding shell and then turn off before it blows away the
ring as well.  Such a coincidence seems unlikely.

\begin{table}
\caption{Summary of results}
\label{tbl:rsltsum}
\begin{center}
\begin{tabular}{rlll}
\tableline
id	& $r_{{\rm ring}}/r_{{\rm HII}}$ & Outer Ring? &
$v_{{\rm shell}}$ \\ \tableline
A & 1.20 & no & 7.2 \\ %r00
B & 1.10 & yes & 7.6 \\ %r01
C & 1.05 & yes & 9.6 \\ %r02
D & 1.21 & no & 7.8 \\  %r03
E & 1.04 & no & 8.7 \\  %r04
F & 1.03 & no & 6.9 \\ %r07
G & 1.04 & yes & 6.8 \\ %r08
XA &  x   & yes & 7.3    \\ %c00
XE &  x   & maybe   & 8.3 \\\tableline\tableline %c04
\end{tabular}
\end{center}
\end{table}

Table~\ref{tbl:rsltsum} summarizes the results of our calculations.
We have chosen to focus on three diagnostic properties: the thickness
of the HII region at the equator (expressed as a ratio of its inner
and outer radii), the presence or absence of a density enhancement
that (if actually azimuthally symmetric) could form an outer ring, and
the expansion velocity of the inner ring.

\section{Discussion}
\label{sec:conclusion}

For the most part the interacting winds scenario seems to be on the
right track for explaining the SN 1987A nebula.  The expansion
velocity of the inner ring is a bit low in these simulations, but it
is close enough that it is reasonable to expect that some combination
of parameters could be found to produce an acceptable match to the
observed value.  

The HII region is the most encouraging development in the simulations.
Although the HII region is somewhat thin in the equatorial plane, the
overall thickness seems to indicate that this problem could be fixed
by judicious choice of wind parameters.  Since the equatorial
thickness of the HII region seems to go down with increasing $\beta$,
further calculations will concentrate on calculations with smaller
$\beta$.

Although we do see mid-latitude clumps in several of our simulations,
we are skeptical that this means a mechanism has been found for
producing the outer rings.  We have seen that it is possible for these
clumps to form even in the nonphotoionizing case, as the neutral shell
breaks up due to Rayleigh-Taylor instabilities.  Consequently, it
seems likely that the clumps would not form axisymmetric rings in
three dimensions. 
However, the edge of the nebular shell created by the ionization front
could have an axisymmetric structure because its position is related
to the density gradient in the RSG wind.
Whether a dense ring could form there depends on the detailed cooling
properties of the gas, which are only approximately accounted for in
our models.

\acknowledgments

This work was supported in part by NASA grant NAG5-8232.

\end{document}